\documentclass[11pt,a4paper]{article}
\usepackage{jheppub}
\usepackage{amssymb}
\usepackage{dsfont}
\usepackage{bibentry}

\newcommand{\bea}{\begin{eqnarray}}
\newcommand{\eea}{\end{eqnarray}}

\newcommand{\ds}{\Delta_{\sigma}}

\title{(No) Bootstrap for the Fractal Ising Model}

\author[a]{John Golden,}
\author[b]{Miguel F. Paulos}

\affiliation[a]{Department of Physics,
Brown University,
Box 1843,
Providence, RI 02912-1843,
USA}
\affiliation[b]{CERN, Theory Division, Geneva, Switzerland}

\abstract{We consider the conformal bootstrap for spacetime dimension $1<d<2$. We determine bounds on operator dimensions and compare our results with various theoretical and numerical models, in particular with resummed $\epsilon$-expansion and Monte Carlo simulations of the Ising model on fractal lattices. The bounds clearly rule out that these models correspond to unitary conformal field theories. We also clarify the $d\to 1$ limit of the conformal bootstrap, showing that bounds can be -- and indeed are -- discontinuous in this limit. This discontinuity implies that for small $\epsilon=d-1$ 
the expected critical exponents for the Ising model are disallowed, and in particular those of the $d-1$ expansion.
Altogether these results strongly suggest that the Ising model universality class cannot be described by a unitary CFT below $d=2$. We argue this also from a bootstrap perspective, by showing that the $2\leq d<4$ Ising ``kink'' splits into two features which grow apart below $d=2$.
}

\begin{document}
\maketitle

\section{Introduction}

The old conformal bootstrap \cite{Ferrara:1974ny,Ferrara:1974pt,Ferrara:1973yt,Ferrara:1971vh,Ferrara:1974nf,Ferrara:1973vz,Polyakov:1974gs} has seen a remarkable revival in recent years. The seminal work \cite{Rattazzi:2008pe} proposed an efficient numerical procedure for extracting information about the space of all conformal field theories, and has been followed by many other works in a variety of contexts \cite{Vichi:2011ux,Vichi-thesis,Rychkov:2009ij,Rattazzi:2008pe,Rattazzi:2010gj,Rattazzi:2010yc,Poland:2011ey,Poland:2010wg,Liendo:2012hy,Kos:2013tga,Gliozzi2013,ElShowk:2012ht,El-Showk:2013nia,ElShowk:2012hu,Beem:2013qxa,Gaiotto2014}. Similarly, the studies \cite{Fitzpatrick:2012yx,Komargodski:2012ek,Alday2013,Alday2014c} have shown that it is even possible to analytically derive completely generic constraints on the spectrum of CFTs. Numerical works have been possible due to increased computer power in the last decades, and rely crucially on an increased understanding of conformal blocks -- see \cite{Dolan2011,Dolan2004,Dolan2001,Poland:2010wg,Fitzpatrick:2013sya,Costa:2011dw,Hogervorst:2013sma,Hogervorst:2013kva,El-Showk2014}.

An important test case for the conformal bootstrap has been the critical 3d Ising model \cite{El-Showk2014,ElShowk:2012ht}, which has been well studied from a variety of theoretical and numerical approaches \cite{Pelissetto2002}. In \cite{El-Showk:2013nia} this study was taken a step further, by considering the critical Ising model in {\em fractional} spacetime dimension $d$, namely for $2<d<4$. Remarkably, the results obtained were consistent with the existence of a CFT living at a corner or kink in the space of unitary theories, whose spectrum very precisely matches that of the Wilson-Fisher fixed point \cite{Wilson:1971dc,Wilson:1972cf,Wilson:1973jj}, \cite{LeGuillou1985,LeGuillou1980}. 
In this work we explore the conformal bootstrap in the range $1<d<2$, extending and completing the work carried out in \cite{El-Showk:2013nia}.

There are multiple reasons why this region is interesting. Firstly, Borel resummation methods \cite{LeGuillou1985} employed to obtain accurate critical exponents for $2<d<4$ become increasingly unreliable for $d<2$. Several alternative techniques have been proposed \cite{LeGuillou1985, Holovatch1993,Bonnier1991,Novotny1992,Wallace1979,Bruce1981} but there is no overall consensus, with large variations in predictions. In contrast, the bootstrap approach has several advantages: although results are numerical in nature, convergence is fast \cite{ElShowk:2012hu}; furthermore, the bootstrap equations are manifestly analytic in $d$ -- it is as difficult to work in integer as fractional dimensions -- providing a definition of the theory for any dimension; finally, the results are non-perturbative in nature and do not depend on any approximation scheme. 

Another reason to consider fractional dimensions is in the context of models on fractal lattices -- for example, one may wish to understand whether the Ising model on such a lattice becomes related to a fractional $d$ field theory in the continuum limit. Initially this seems unlikely, as generic fractal spaces are expected to break translation invariance even in the continuum limit. Nevertheless, such systems do show critical behaviour at finite temperature, and so there have been multiple attempts to compare the Wilson-Fisher critical exponents with those of fractals, with mixed results~\cite{monceau1998magnetic}. Furthermore, it has been suggested that certain fractals \cite{Gefen1984,Gefen1984a,Gefen1980} (those with small lacunarity) do recover translational invariance in the continuum. 

A final reason to consider $1<d<2$ is to understand the limit $d\to 1$. From a theoretical standpoint,  this is an interesting limit because the number of independent conformal cross-ratios of four point functions jumps discontinuously from two to one when $d=1$. Furthermore, for $d=1$ there are only spin-0 conformal primaries whereas for any $d>1$ there is an infinite set of spin-$L$ representations. We show that in this limit the crossing equations decompose into two sectors, only one of which is present at exactly $d=1$. This implies that the bootstrap results can be discontinuous in the $d\to 1$ limit, and remarkably this is precisely what we find. From a practical standpoint, we shall be able to compare our results with those of the $d=1+\epsilon$ expansion \cite{Wallace1979,Bruce1981} -- these are models of interfaces between two phases which can show critical behaviour and have been argued to describe the Ising model. In any case, they lead to concrete analytic predictions for critical exponents which we are able to compare with our numerical results.

Although our approach is essentially that of \cite{El-Showk:2013nia}, our results are very different. Our bounds not only unequivocally rule out essentially {\em all} theoretical predictions for the critical Ising model in these dimensions, they also exclude the critical points found on Monte Carlo lattice simulations. To be clear, by this we mean that all these models and predictions cannot possibly describe {\em unitary, conformal} field theories. This is not really a big issue for the Monte Carlo simulations, since one does not necessarily expect the fixed points to have conformal symmetry; but it is quite striking that agreement with theory fails so catastrophically given the beautiful agreement found for $d>2$. We will argue that this is evidence for a qualitative difference in the nature of the Ising model universality class below $d=2$, and will back this up with a detailed analysis of the spectra of solutions to crossing symmetry in the neighbourhood of the hypothetical Ising model for $d\lesssim 2$.

This paper is organized as follows. After a brief review of the conformal bootstrap program, in section~\ref{mainsection} we present our bounds and then compare them with theoretical predictions for the Ising critical exponents. In section~\ref{sec:fractals} we give a short review of fractal lattice models and compare our bounds with various results in the literature. We then carefully study the $d\to 1$ limit in section~\ref{sec:dto1}. We show that in this limit there can be a discontinuity in the bounds, a discontinuity which we find numerically. This discontinuity rules out the predictions of the $\epsilon=d-1$ expansion, and indeed of any model which predicts that in this limit one should have $(\Delta_\sigma,\Delta_\varepsilon)\to (0,1)$, or equivalently $(\eta,\nu)\to (1,\infty)$. We finish with a discussion.

\section{Review}\label{sec:review}

In this section we briefly review the numerical bootstrap program and how it is used to constrain the space of unitary conformal field theories. We will not be too detailed, and refer to reader to {\em e.g.} \cite{El-Showk2014} for more details.  We begin with the four-point function of a scalar field $\sigma$, 
\begin{eqnarray}
\langle \sigma(x_1)\sigma(x_2)\sigma(x_3)\sigma(x_4)\rangle&=&\frac {g(u,v)}{x_{12}^{2\Delta_\sigma}x_{34}^{2\Delta_\sigma}}.
\end{eqnarray}
Conformal symmetry forces this particular kinematic dependence. In particular, the function $g(u,v)$ can only depend only on the conformally invariant cross-ratios
\begin{eqnarray}
u=\frac{x_{12}^2 x_{34}^2}{x_{13}^2\,x_{24}^2},\qquad v=\frac{x_{14}^2 x_{23}^2}{x_{13}^2\,x_{24}^2}.
\end{eqnarray}
In a conformal field theory there is also an (exponentially fast  \cite{Pappadopulo:2012jk}) convergent operator product expansion (OPE), which gives us dynamic constraints on the correlator. For instance, equality of the OPE expansions in the (12) and (14) channels forces
\begin{eqnarray}
g(u,v)=\sum_{\mathcal O} \lambda_{\sigma \sigma \mathcal O}^2 G_{\Delta,L}(u,v)&=&\left(\frac{u}{v}\right)^{\Delta_\sigma}\sum_{\mathcal O} \lambda_{\sigma \sigma \mathcal O}^2 G_{\Delta,L}(v,u).
\end{eqnarray}
The conformal block $G_{\Delta,L}(u,v)$ represents the contribution to the four point function from a primary operator $\mathcal O$ with conformal dimension $\Delta$ and traceless-symmetric spin $L$, together with all its descendants. Such operators can appear in the $\sigma \times \sigma$ OPE with coefficient $\lambda_{\sigma \sigma \mathcal O}$. This equality can be rewritten in an obvious way as a linear equation with positive coefficients,
\begin{eqnarray}
\sum_{\mathcal O} \lambda_{\sigma \sigma \mathcal O}^2 F^{\Delta_\sigma}_{\Delta,L}(u,v)=0. \qquad 
\end{eqnarray}
The idea now is to think of the above as an abstract equation, involving a continuously infinite set of constraints on the  continuously infinite set of parameters $\lambda_{\Delta,L}$, which {\em any} CFT must satisfy. In a typical CFT the actual operators appearing in a given four point function are countably infinite, but in analysing this equation we will not assume it. 

Unitarity guarantees positivity of the squares of the OPE coefficients, which in turn guarantees the non-triviality of the equation above -- for even if the ensemble of functions $F_{\Delta,L}^{\Delta_\sigma}(u,v)$ for all $\Delta,L$ would form a ``basis'', it might not be possible to find a solution for this equation with {\em positive} coefficients. Of course, one trivial solution would be to set all coefficients to zero -- but this is impossible due to the guaranteed presence of the identity operator which appears with unit coefficient.

To proceed we discretise the set of constraints, typically by Taylor-expanding the functions $F$ up to some finite order. We also impose a cut-off on the allowed spins and conformal dimensions. In practice one checks that the results are independent of the choice of cut-offs if we choose them sufficiently high. The larger the set of constraints, the higher the cut-off one must take. 
 With these modifications we can now solve the problem -- linear equations with linear inequalities are examples of linear programs for which efficient algorithms have been developed since the 1950's.  In this work we use a variation of Dantzig's simplex method~\cite{dantzig1955generalized} suitable for a continuous set of $\Delta$. The numerical package used has been developed in the {\tt Julia} language\footnote{The package, {\tt JuliBootS} will be officially released soon but is already available on GitHub: {\tt http://github.com/mfpaulos/JuliBoots}.}, but the methods are otherwise essentially those of \cite{El-Showk2014}, and we will follow their conventions for the truncation of the problem\footnote{In particular, truncations are labelled by a number $n_{max}$ which determines the size of the truncation.}.

Once we have a method for solving the equations, we can derive various constraints on the spectrum of operators appearing in the sum rule. Here we will be concerned with deriving the maximal possible gap in the scalar sector -- in other words, the maximal allowed dimension for the first scalar operator appearing in the $\sigma \times \sigma$ OPE, which we shall call $\varepsilon$:
\begin{eqnarray}
\sigma \times \sigma \simeq \mathds 1+\varepsilon+\ldots
\end{eqnarray}
In order to achieve this, we remove scalar operators from the sum rule up to some value $\Delta_{\varepsilon}$ until we can no longer find a solution. The precise value depends on the number of constraints, but it is guaranteed to decrease as this number is increased. In this way, for any truncation of the equations we derive a perfectly valid upper bound on the dimension of the leading scalar. This bound also constrains the allowed ranges of the critical exponents $\eta$ and $\nu$ of the Ising model through the relations
\begin{equation}\label{eq:exponent_operator_relations}
    \Delta_{\varepsilon}=d-\frac{1}\nu, \qquad
	\Delta_{\sigma}=\frac 12 (d-2+\eta).
\end{equation}

We can repeat this procedure for several values of $\Delta_\sigma$, thereby obtaining a bound curve. This is what we shall do in the following sections, varying the spacetime dimension as we go along. This is sufficient to rule out large regions in the parameter space of conformal field theories. However, we can go further. If we place ourselves precisely at the boundary between allowed and disallowed theories, we can extract a unique solution to crossing symmetry \cite{ElShowk:2012hu} -- that is, the low-lying spectrum of a hypothetical CFT living at this boundary. As we move along the boundary, these spectra can behave in interesting ways, displaying sharp rearrangements \cite{El-Showk2014} which can signal interesting theories -- in particular, this is the case for the Wilson-Fisher fixed point in $2<d<4$. In this note we will determine spectra and use them as a guide to a better understanding of the nature of these theories.

\section{Bounds for $1<d<2$}
\label{mainsection}

\begin{figure}[h]
	\centering
		\includegraphics[width=15cm]{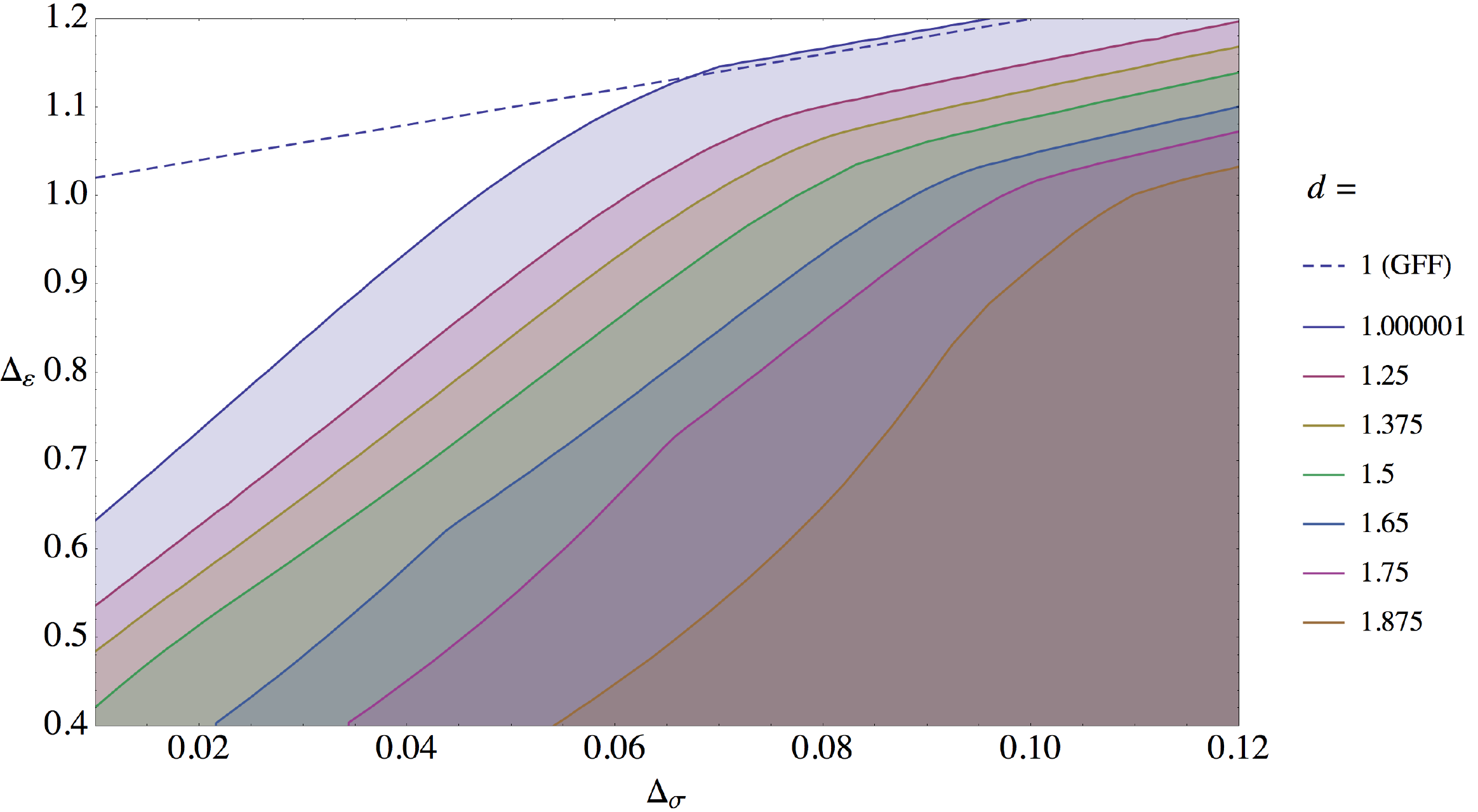}
		\caption{Bounds for $1<d<2$. In the terminology of \cite{ElShowk:2012ht} these were done with $n_{max}=15$. They correspond to a truncation of the constraints to 136 components.}
	\label{fig:allbounds}
\end{figure}

Our numerical results are presented in figure~\ref{fig:allbounds}. It shows upper bounds on the conformal dimension of the leading scalar primary $\varepsilon$ as a function of $\Delta_\sigma$  for various values of spacetime dimension $d$. 
These bounds provide universal constraints on the space of unitary CFTs in $1<d<2$. As a consistency check the generalized free scalar, which is described by the curve $\Delta_{\epsilon}=2\Delta_{\sigma}$ is well below our bounds in any $d$; we have also checked that for $d=2$ we reproduce existing results in the literature \cite{Rychkov:2009ij}. Notice however that the bound for $d=1.00001$ does {\em not} match the result in \cite{Gaiotto2014} for $d=1$. Indeed for $d=1$ there is a generalized free fermion (GFF) solution available which has $\Delta_{\varepsilon}=1+2\Delta_\sigma$, shown as a dashed line in our figure. We will have more to say about this in section \ref{sec:dto1}.

\begin{figure}[h]
	\centering
 	\includegraphics[width=17cm]{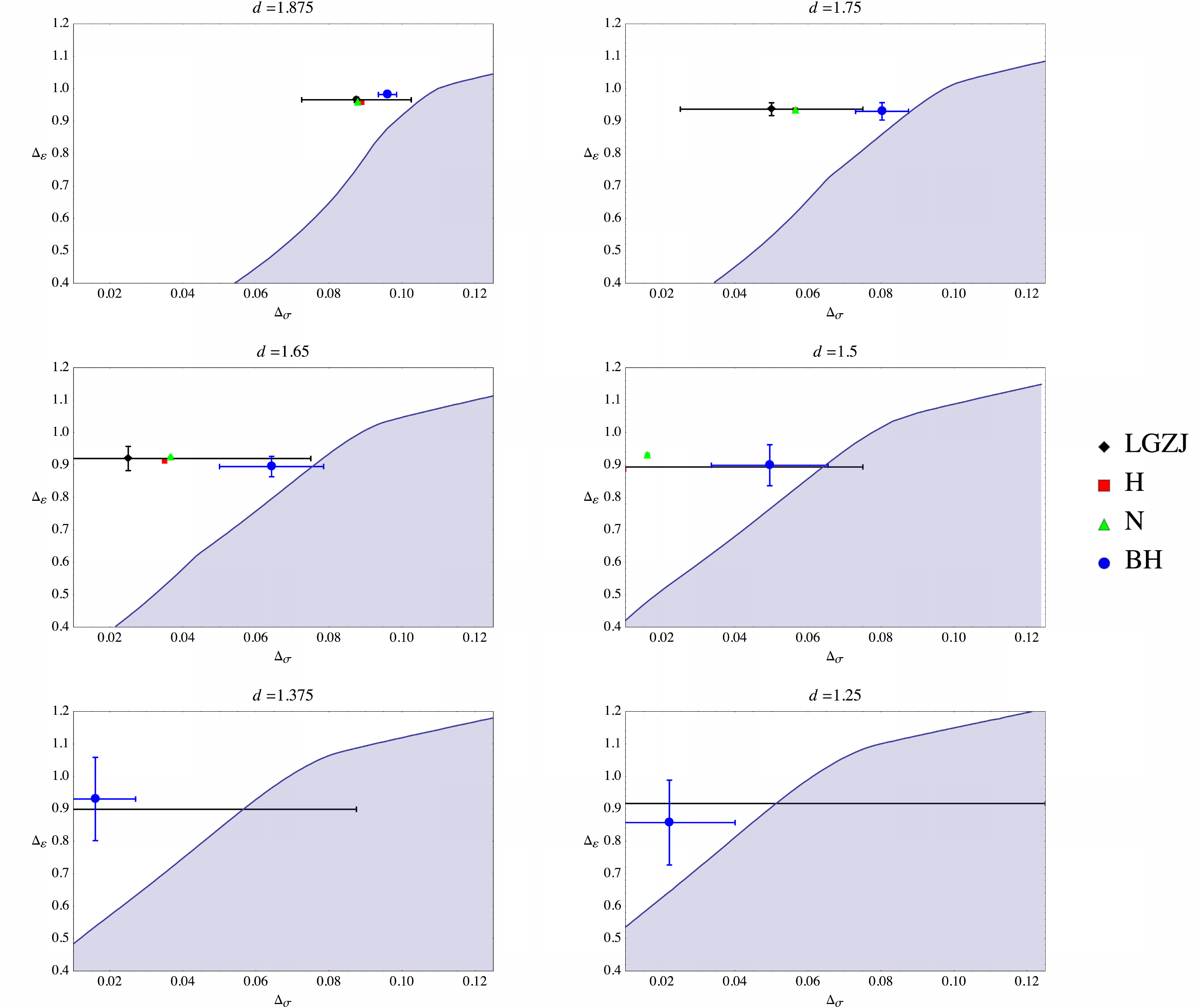}
	\caption{Comparison with different theoretical approaches, taken from \cite{Holovatch1993}. For small enough $d$ some of the predictions lie outside the unitarity bounds and we do not show them.}
	\label{fig:experiment}
\end{figure}

For now we would like to see how these bounds relate to various predictions for the Ising model critical exponents in fractional dimension. Here we shall mention only four, which were conveniently compiled in \cite{Holovatch1993}:
\begin{itemize}
\item{\bf LGZJ} -- Le Guillou and Zinn-Justin \cite{LeGuillou1985} used Borel resummation methods on 4-loop ($\epsilon^5$) epsilon expansion \cite{Wilson1972a} results to obtain accurate critical exponents for $\epsilon=1$ and $\epsilon=2$. In later work \cite{LeGuillou1987} the critical exponents were obtained for various values of dimension between one and four. 
\item{\bf H} -- One can also compute the critical exponents in $\phi^4$ theory directly in the desired dimension, as proposed in \cite{Parisi1993} and applied with success to $d=3$ in \cite{LeGuillou1977,LeGuillou1980}. Of more interest to us are similar computations performed by Holovatch \cite{Holovatch1993} for various values $1<d<3$ to three loops. 
\item{\bf N} -- A different approach is an interpolation method for numerical transfer matrix data of Novotny \cite{Novotny1992}. The idea there is to rewrite the Ising model lattice partition function in terms of a suitable transfer matrix, in such a way that computations can be generalized for arbitrary dimensionality. Critical exponents are then determined by finite-size scaling.
\item{\bf BH} -- Bonnier and Hontebeyrie \cite{Bonnier1991} considered the Ising model in a static magnetic field with a variational parameter defined as $h = H + \lambda(1-t)$, with all final results evaluated at $t=1$. They then approximated the functional form of the critical exponents by Pad\'e approximants and tuned $\lambda$ to find coherent values. 
\end{itemize}
Figure~\ref{fig:experiment} compares our bounds with these approaches for several values of $d$. Remarkably, our bounds essentially completely rule out these four different sets of predictions. These methods are usually tuned to agree with the exact 2d Ising results. Accordingly they are more or less consistent with each other for larger $d$, and rapidly develop large error bars and relative disagreements as $d$ tends to one. But even if we focus on the first plot with, where we are still relatively close to $d=2$, we see that our bounds already exclude quite clearly all four different approaches. 
\pagebreak

The Borel-resummed $\epsilon$-expansion results are particularly interesting. In previous work it was found that the predictions of this method are beautifully consistent with those of the bootstrap \cite{El-Showk:2013nia}. In that work, bounds were computed for various dimensions $2<d<4$. All such bounds had sharp kinks located very precisely at the location predicted by the $\epsilon$-expansion computations. Here however the situation is very different: the predictions are simply not consistent with the bootstrap results, even for $d$ as large as $1.875$. This suggests that the nature of the Wilson-Fisher fixed point is different for $d<2$. In particular, whatever it is, it seems that it cannot be a unitary conformal field theory.

If this is really so, we would like to see some evidence for it directly from a bootstrap perspective. For $d>2$, the Wilson-Fisher fixed point lies on a kink. It is interesting then to consider what happens to this kink for $d<2$. However, we immediately run into a puzzle: a closer look at our curves seems to show that there is not one but two inflection points! They can be seen especially clearly in the bounds for dimensions $1.875$ and $1.65$, and seem to fade away below $d=1.5$. To clear up the situation we must examine the spectra of the solutions to crossing symmetry along the boundary of the bounds. Kinks in the bounds have previously \cite{El-Showk2014} been shown to be related to rearrangements in the spectrum of solutions as we vary $\Delta_\sigma$. Therefore, instead of looking for features in the bound plots, we shall instead consider the spectra and look for such rearrangements there.

\begin{figure}[h]
	\centering
	\begin{tabular}{cc}
		\includegraphics[width=7cm]{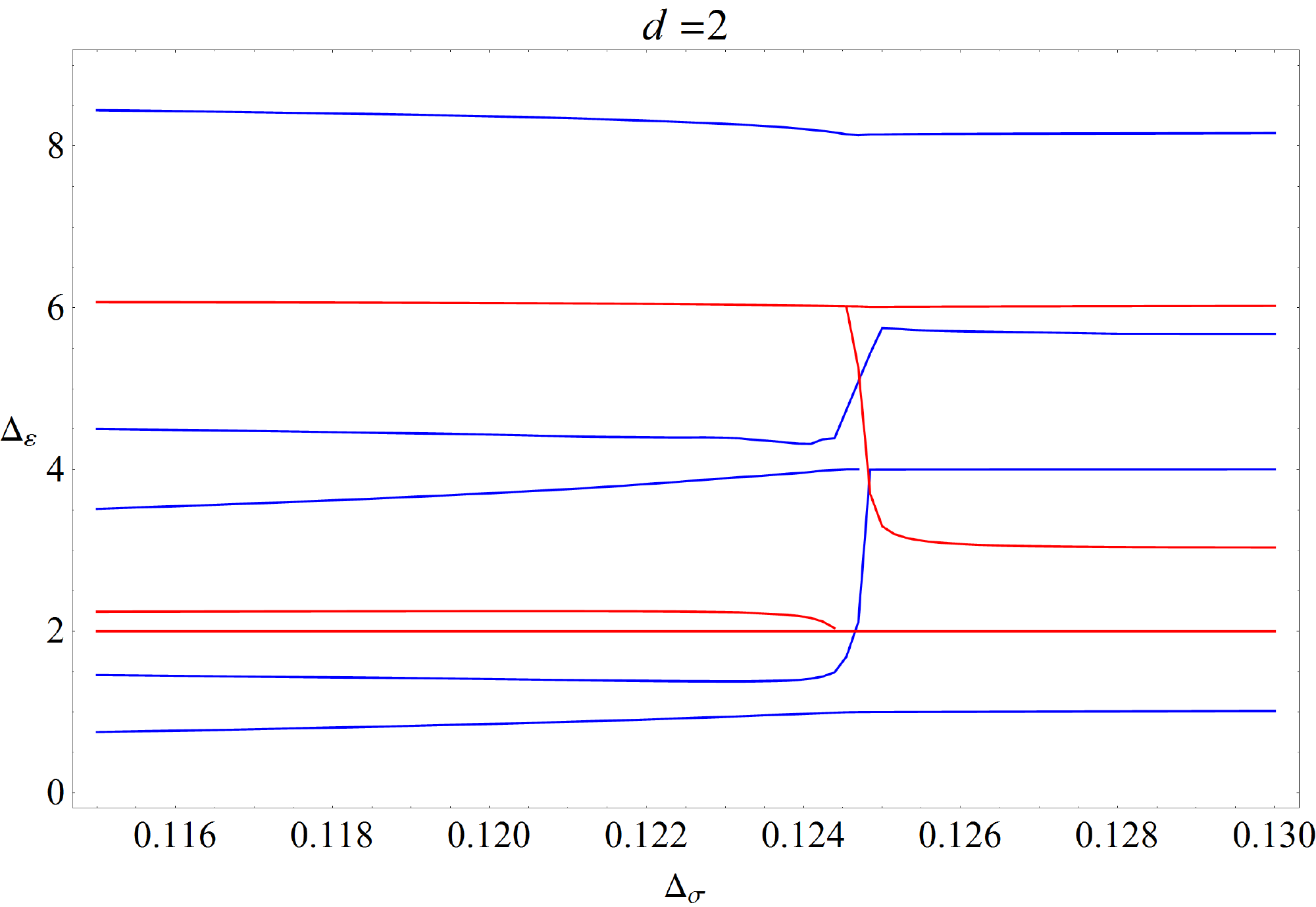} &
		\includegraphics[width=7cm]{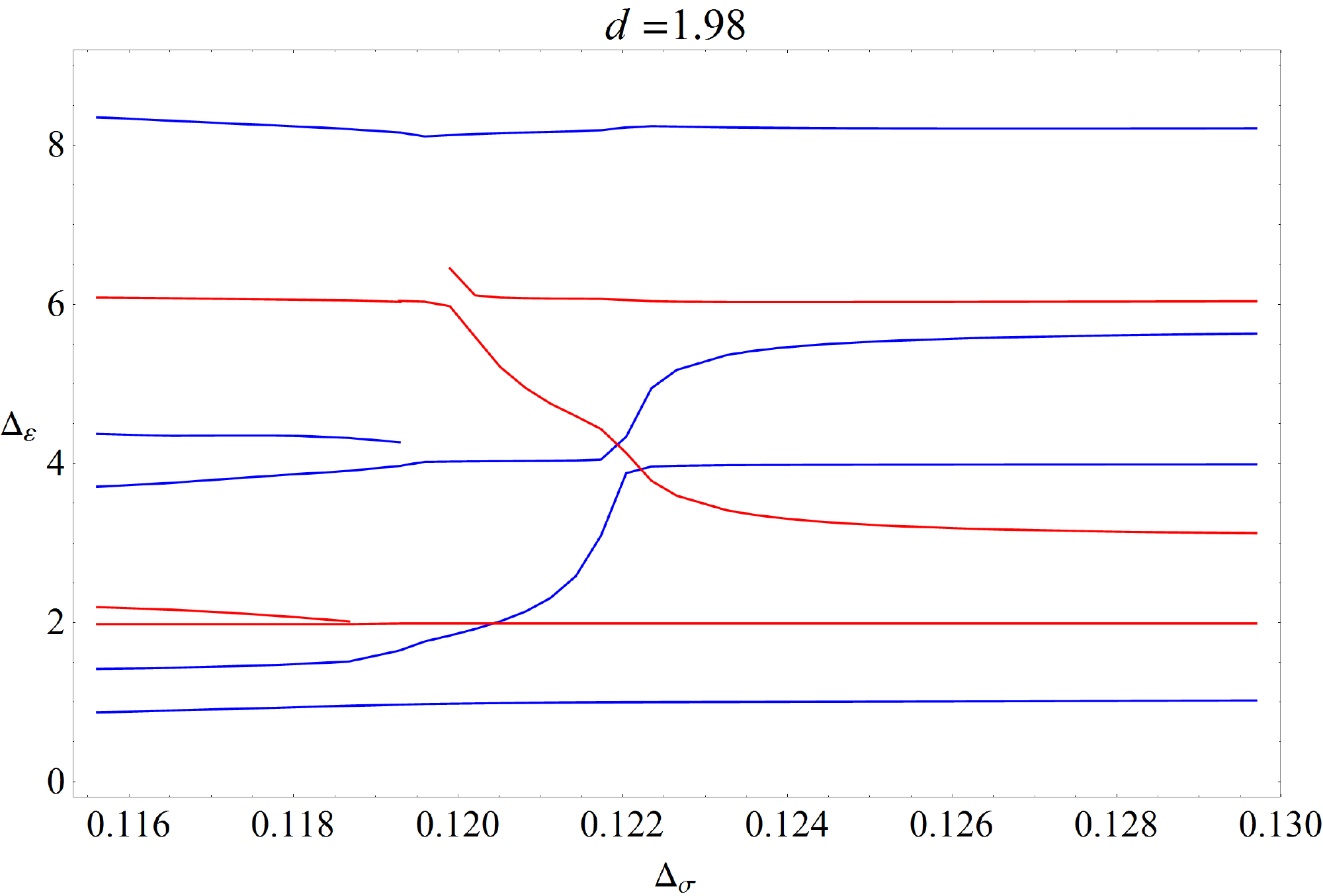}
	\end{tabular}
	\caption{Spectra of solutions to crossing symmetry along the edge of the allowed region. Shown are the low-lying spin-0 and spin-2 operators. Plots were made using truncations to 78 components ($n_{max}=11$).}
	\label{fig:specSmall}
\end{figure}

We begin by considering what happens as $d$ is lowered below two. In figure \ref{fig:specSmall} we show the spectra close to $\Delta_\sigma=1/8$ (which is the correct value for the critical Ising model in $d=2$) for $d=1.98$ and $d=2$. For $d=2$ we see that there are sharp operator rearrangements taking place in both the spin 0 and spin 2 sectors. These two rearrangements lie very close to each other, and indeed we have checked that as we increase the level of the truncation ({\em i.e.} as our bounds become stronger) they approach each other. For $d=1.98$ we see that these rearrangements are still present, but a significantly larger distance apart.
Going down in dimension, we display wider-range spectra for $d=1.8$ and $d=1.5$. In the first the two features are again clearly seen, but now a very wide distance apart. By the time we get down to $d=1.5$ one them seems to have disappeared, or at the very least moved out towards $\Delta_{\sigma}=0$.
\begin{figure}[h]
	\centering
	\begin{tabular}{cc}
		\includegraphics[width=7cm]{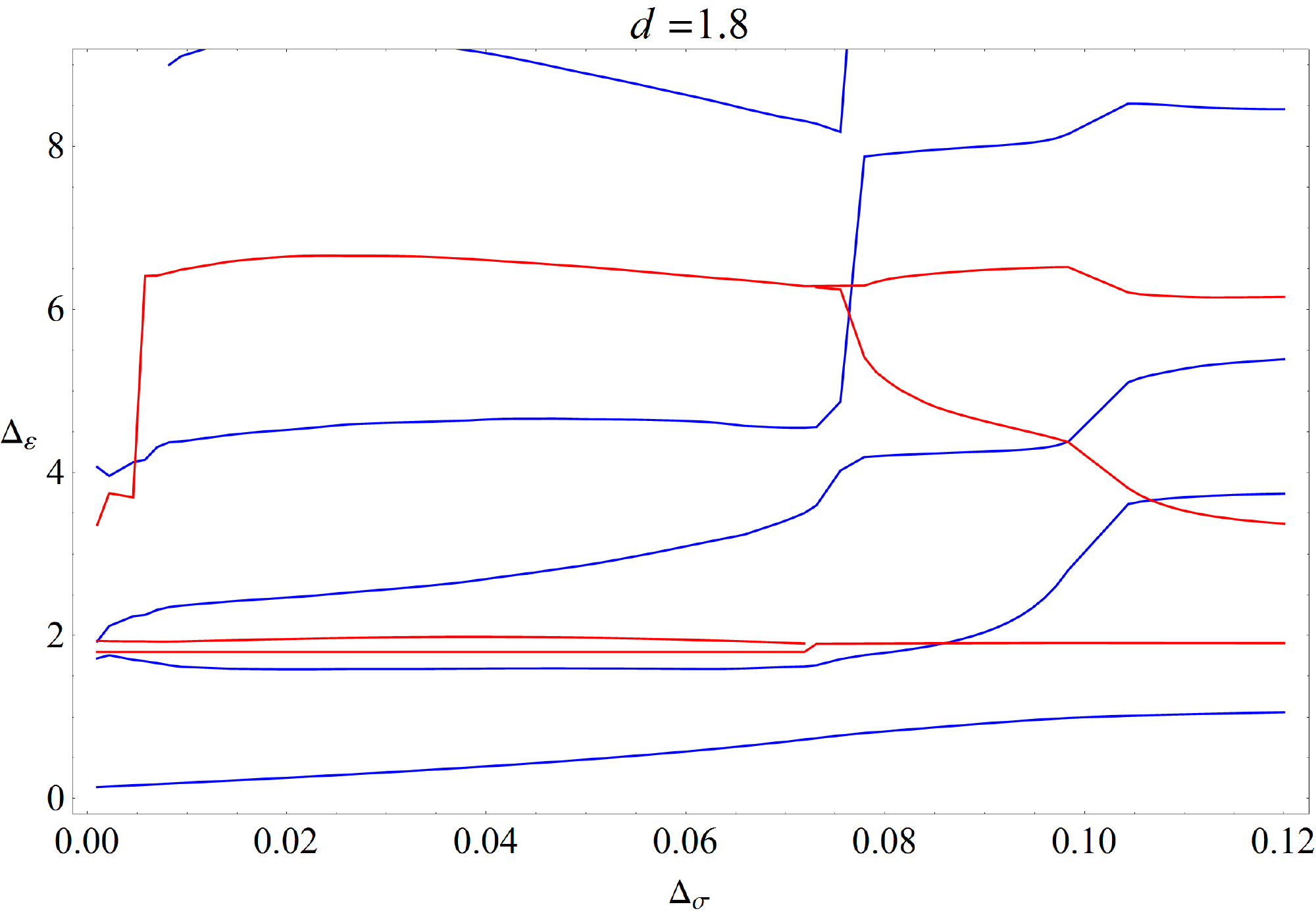} & 
		\includegraphics[width=7cm]{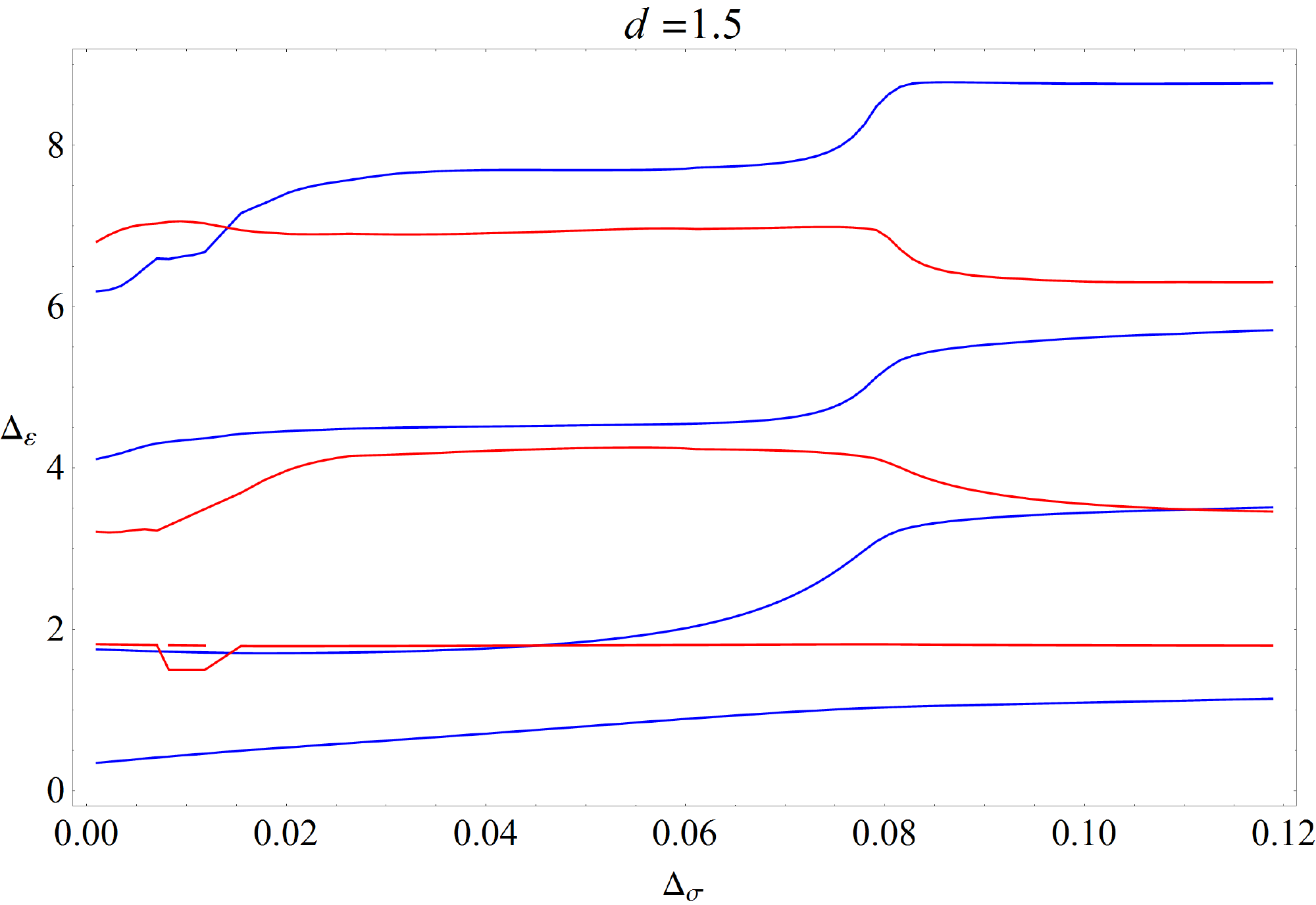} 
	\end{tabular} 
	\caption{Spectrum along the bound curves for dimensions 1.8 and 1.5. Two spectra rearrangements can be clearly seen in the first case. For $d=1.5$ the first one seems to have disappeared or moved out towards the origin.}
	\label{fig:spectra}
\end{figure}
The tentative conclusion seems to be that the unique kink observed in $2\leq d<4$ actually splits into two distinct features which grow apart as $d$ is lowered: one of the kinks moves towards $\Delta_\sigma=0$ while the other one eventually becomes the feature seen in the bound plots for $d\simeq 1$. From the perspective of the bootstrap program, this qualitative difference in the bounds (and the set of solutions of crossing that follow from them) provides strong evidence that the nature of the Wilson-Fisher fixed point drastically changes below $d=2$.

\section{The Ising model on fractal surfaces}\label{sec:fractals}

The results discussed so far have a somewhat formal character since it is not clear to what (if any) systems they could be applied. However there is a very natural guess -- since we are considering theories in fractional spacetime dimension, we may attemp to model this by considering a system living on a fractal lattice, whose dimension in the continuum limit becomes non-integer. This was precisely the approach first attempted by Mandelbrot and collaborators \cite{Gefen1980}, who considered the Ising model on various kinds of fractal lattices. An incomplete list of later work includes \cite{Bab2005,Bab2009,Bonnier1986,Bonnier1987a,Bonnier1989,Carmona1998}. These works show that for large classes of fractals it is possible to find critical points and associated critical exponents.

There are several issues which arise when we attempt to identify a particular theoretical result in fractional dimension with a concrete setup of the Ising model on a fractal lattice. Firstly, it is generally assumed that close to criticality we obtain not only scaling invariance, but also full translational symmetry. For fractal structures this does not happen, since there are voids in the lattice which persist at all length scales. Furthermore, fractals are defined by several parameters other than dimensionality -- {\em e.g.} ramification order and lacunarity\footnote{The ramification order is the number of bonds that must be cut to isolate an arbitrarily large sub-lattice, while lacunarity is related to the mean square deviation of the mass of a fractal contained in a shell of fixed radius.} -- features which again are present at all length scales, surviving the continuum limit. Critical exponents (or even the existence of a critical point) depend on these -- a phase transition can occur only for infinite ramification order \cite{Gefen1980}, and it has been claimed that translation invariance can be recovered in the limit of zero lacunarity \cite{Gefen1983}. Unfortunately, fractal lattices with low lacunarity are more difficult to model numerically as they have more sites.

Given this restriction to fractals with low lacunarity and infinite ramification order, all of the in-depth numerical studies in the literature focus on Sierpinski carpets. The Sierpinski carpet $SC(b,c)$ is constructed by dividing a square into $b^2$ subsquares and then removing $c^2$ of them; this process is then iterated infinitely many times on the remaining squares to achieve a mathematical fractal. In practice, Monte Carlo simulations are run on fractals where the segmentation and deletion procedure has been iterated a finite number $k$ times. The fractal $SC(b,c)$ is understood to represent a surface with fractional dimension, known as the Hausdorff dimension $d_H = \log(b^2 - c^2)/\log(b)$. 

When modelling the Ising model on a fractal lattice, there are several choices to be made which can significantly affect the final results. Firstly, in the construction of the fractal itself there are numerous schemes for choosing which $c^2$ subsquares to delete. We will mention only two: in $SC_a$ fractals, the subsquares are deleted from the center of the larger square, while in $SC_b$ the squares are deleted in an alternating fashion starting from the uppermost left corner. $SC_a(x,y)$ and $SC_b(x,y)$ have identical $d_H$ but different lacunarities. 
Secondly, the numerical simulation itself depends on several parameters: the number of iterations $k$ to include, the boundary conditions satisfied by the lattice (i.e. periodic vs. free), whether the spins are placed in the center of the squares or at vertices, and the method of calculation chosen (Monte Carlo, real-space renormalization, high-$T$ expansions, etc.). For a thorough review of these intricacies, see~\cite{monceau1998magnetic}. 

For the purposes of this (admittedly limited) review of this topic, these different choices will play no major role. We have chosen to present a selection of fractals from the relatively recent work of Bab, Fabricius, and Albano~\cite{Bab2009}. They studied a range of Sierpinski carpets using Monte Carlo simulations with the Metropolis algorithm, periodic boundary conditions, and spins placed at lattice vertices. More information about their approach can be found in~\cite{Bab2005,PhysRevE.74.041123}. Other results are available~\cite{monceau1998magnetic,Carmona1998} but differ from the results of~\cite{Bab2009} by comparitively minor amounts and do not affect the qualitative conclusions of the comparison with our bounds. 

One important question raised by this and other works is that of the ``true'' dimensionality of the system. Specifically, does the Hausdorff dimension satisfy the known hyperscaling relationship for small dimensions ($d=2\beta/\nu +2 - \eta$)? For certain fractals the critical exponents lead to an `effective' dimmension $d_{eff}$, given by the hyperscaling relationship, that roughly matches $d_H$. However, in general the two dimensions are quite different, suggesting that $d_H$ is perhaps not the relevant dimension for these critical systems (though a rigorous understanding of this phenomenon is lacking). 

\begin{figure}[ht]
	\centering
		\includegraphics[width=13cm]{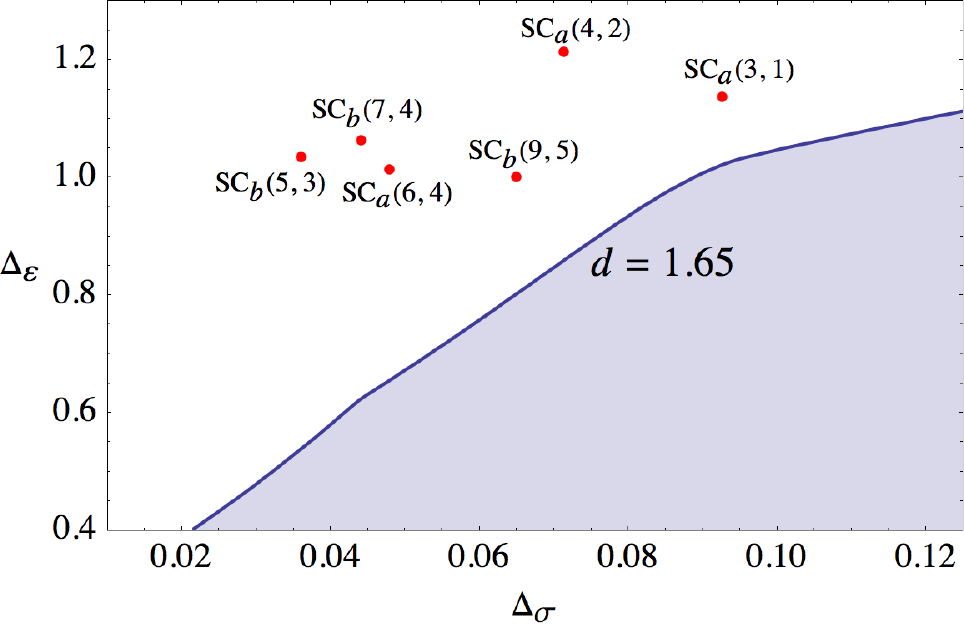} 
		\caption{Comparison of fractal Ising models on Sierpinski carpets with the $d=1.65$ bound.} %Detailed information about each fractal can be found in table~\ref{table:fractals}.}
	\label{fig:fractal_ising}
\end{figure}

We make a comparison between our bounds and the results of~\cite{Bab2009} in figure~\ref{fig:fractal_ising}. To convert their data into conformal dimensions we have used eq.~\ref{eq:exponent_operator_relations} together with the hyperscaling relation, assuming that $d=d_{eff}$. Using $d_H$ instead of $d_{eff}$ does not move any of the points inside our bounds, and in fact it can lead to negative values for $\Delta_\sigma$. In doing the comparison we have shown a single one of our bounds, namely that for dimension $d=1.65$. We are allowed to do this because the bounds become more constraining for higher $d$ and $d_{eff}>1.65$ for all theories we considered. We have included only a representative selection of fractal Ising models, but we have checked that  other results in literature~\cite{monceau1998magnetic,Carmona1998} fall within this general area and none satisfy our bounds.  

Unfortunately our results also shed little light on the question of whether decreasing lacunarity corresponds to a theory closer to the fractional $d$ field theory in the continuum limit. For example, of the fractals considered in fig.~\ref{fig:fractal_ising}, $SC_a(3,1)$ has the highest lacunarity while $SC_b(9,5)$ has the lowest, but they both appear roughly the same distance from the bound. It is difficult to draw any firm conclusions from this analysis due to the low number of numerical results in the literature. 
%

% removing this table for the time being since I don't know that it really added too much interesting information 
%\\
%\begin{table}[h]\label{table:fractals}
%\begin{center}
%\begin{tabular}{|l|lllllll|}\hline
%Fractal & $d_H$ & $d_{eff}$ & $\gamma$ & $\nu$ & $\Delta_\sigma$ & $\Delta_\epsilon$ & $k$\\ \hline
% $\text{SC}_a(3,1)$ & 1.893 & 1.864 & 2.22 & 1.323 & 0.0926 &
%   1.13 & 6\\ 
% $\text{SC}_a(4,2)$ & 1.792 & 1.947 & 3.1 & 1.729 & 0.0712 &
%   1.21 & 6\\ 
% $\text{SC}_a(6,4)$ & 1.672 & 2.03 & 2.9 & 1.52 & 0.0478 &
%   1.01 & 4\\ 
% $\text{SC}_b(5,3)$ & 1.723 & 1.797 & 2.5 & 1.455 & 0.0359 &
%   1.03 & 4\\ 
% $\text{SC}_b(7,4)$ & 1.797 & 2.01 & 2.62 & 1.364 & 0.044 &
%   1.06 & 4\\ 
% $\text{SC}_b(9,5)$ & 1.832 & 1.947 & 2.18 & 1.204 & 0.0648 &
%   1.001 & 4\\ \hline
%\end{tabular}
%\caption{A selection of fractals from~\cite{Bab2009}. \commentjg{is this table necessary?}}
%\end{center}
%\end{table}
%

\section{The limit $d\to 1$}\label{sec:dto1}

\subsection{Bootstrap equations}

Conformal symmetry of correlation functions effectively reduces their dependence to a handful of cross-ratios, which are invariants under the action of the conformal group. The number of independent cross-ratios depends on the number of fields and is naively given by $n(n-3)/2$. However, this is not strictly true since cross-ratios are built out of finite-dimensional coordinate vectors, and this can lead to non-linear identities between them. In particular, for four-point functions the two cross-ratios $u$ and $v$ are not independent in $d=1$, since
\bea
1-\frac{(x_1-x_2)(x_3-x_4)}{(x_1-x_3)(x_2-x_4)}-\frac{(x_1-x_4)(x_2-x_3)}{(x_1-x_3)(x_2-x_4)}=0 \Rightarrow v=(1-\sqrt{u})^2
\eea
Defining $u=z \bar z, v=(1-z)(1-\bar z)$ ($z,\bar z$ are complex conjugates in the Euclidean domain), this equality corresponds to restricting oneself to $z=\bar z$. In particular this implies that correlation functions and conformal blocks in one dimension can only depend on a single parameter. On the other hand, we do know that there exist solutions to crossing symmetry in $d$ arbitrarily close to one, which will necessarily depend on two cross-ratios -- generalized free fields provide an example -- so it seems like there must be some discontinuity involved in the $d\to 1$ limit.

To examine this question, we consider conformal blocks as functions of spacetime dimension. Conformal blocks are eigenfunctions of the Casimir operators of the conformal group \cite{Dolan2011}. In terms of the $z,\bar z$ variables the action of the quadratic Casimir takes the form~\cite{Hogervorst2013a}:
\begin{eqnarray}
\mathcal C_2^{(\varepsilon)}=D_z+D_{\bar z}+2 \varepsilon\, \frac{z \bar z}{z-\bar z}\left[(1-z) \partial_z-(1-\bar z)\partial_{\bar z}\right],\\
D_z\equiv (1-z) z^2\, \partial_z^2-(a+b+1)\, z^2 \partial_z-a b z, \qquad \varepsilon=\frac{d-2}2.
\end{eqnarray}
with $a=\Delta_1-\Delta_2$, $b=\Delta_3-\Delta_4$ are related to the dimensions of the operators in a four point function. The conformal blocks then satisfy the equation:
\begin{eqnarray}
(C_2^{\varepsilon}-c_2) G_{\Delta,L}(z,\bar z)=0,\qquad
c_2=\frac 12 \left[ L(L+2\varepsilon)+\Delta(\Delta-2-2\varepsilon)\right].
\end{eqnarray}
Since the $d\to 1$ limit is related to $z\to \bar z$, let us expand the blocks around $z=\bar z$. Defining $x,y=(z\pm \bar z)/2$, we have
\begin{eqnarray}
G_{\Delta,L}(z,\bar z)=g_{\Delta,L}(x)+y^2 h_{\Delta,L}(x)+\mathcal O(y^4)
\end{eqnarray}
It turns out that the function $g_{\Delta,L}(x)$ satisfies a differential equation which can be obtained by considering the action of the quartic Casimir together with the quadratic one~\cite{Hogervorst2013a}. The exact form of this equation is not important here -- it is sufficient to mention that for $L=0$ the block satisfies a third-order differential equation for general $d$, whereas for $L>0$ it satisfies a fourth-order equation. On the other hand, the Casimir equation above implies:
\begin{eqnarray}
\tilde D_{\varepsilon} g_{\tiny{\Delta,L}}(x)=-(1+2\varepsilon)\, (1-x)x^2\, h_{\Delta,L}(x), \\
\tilde D_{\varepsilon}\equiv \frac 12(1-x)x^2\, \partial_x^2-(1+a+b+\varepsilon)\,x^2\,\partial_x-2\,ab\,x-c_2
\end{eqnarray}
We are interested in the limit $d\to 1 \Rightarrow \varepsilon \to -\frac 12$. In this limit there are two distinct possibilities. Suppose first that $h_{\Delta,L}$ is finite in this limit. Then it follows that the conformal block at $z=\bar z$ satisfies a second order differential equation, namely $\tilde D_{-\frac 12}\, g_{\Delta,L}=0$. Now, we already know that this function generically satisfies a third or fourth order differential equation, so this can only be true if such an equation factorizes for $\varepsilon=-\frac 12$. We find this to be precisely so only for the cases $L=0$ and $L=1$. We have exact expressions for these blocks when $z=\bar z$, derived in~\cite{ElShowk:2012ht}:
\begin{gather}
G_{\Delta,0}(z) =
\left(\frac{z^{2}}{1-z}\right)^{\Delta/2}
\, _3F_2\left({\textstyle\frac{\Delta}2,\frac{\Delta}2,\frac{\Delta
   }{2}-\varepsilon ;\frac{\Delta+1 }{2},\Delta -\varepsilon}
   ;\frac{z^2}{4 (z-1)}\right)\,,  
\label{eq:3F2-0}  \\
G_{\Delta,1}(z) =\frac{2-z }{2 z} \left(\frac{z^2}{1-z}\right)^{\frac{\Delta +1}{2}}
\, _3F_2\left({\textstyle\frac{\Delta+1 }{2},\frac{\Delta+1
   }{2},\frac{\Delta+1 }{2}-\varepsilon ;\frac{\Delta
   }{2}+1,\Delta -\varepsilon };\frac{z^2}{4 (z-1)}\right)\,.
   \label{eq:3F2-1} 
\end{gather}
Taking $d\to 1$, or equivalently $\varepsilon \to -1/2$ we get:
\begin{eqnarray}
\lim_{d\to 1}\, G_{\Delta,0}(z)=\,\lim_{d\to 1} G_{\Delta,1}(z)= z^\Delta \ _2 F_1(\Delta,\Delta,2\Delta;z)
\end{eqnarray}
so they are the same. A cross-check is that the values of the Casimir for $L=0$ and $L=1$ are the same when $d=1$. Notice however that for $d$ infinitesimally close to one these blocks have a smooth analytic continuation into the region $z\neq \bar z$ where they {\em are} distinct.

For higher spins instead $h_{\Delta,L}(x)$ diverges as $d\to 1$. Considering higher orders in the $y$ expansion, no new sources of divergences are introduced, so that all other terms are either finite or divergent as $1/(d-1)$. This divergence suggests we should change the normalization of the $L\geq 2$ conformal blocks by the same factor - or equivalently, absorbing it into the OPE coefficients. When we do so, we see that the purely $z=\bar z$ piece of the block decouples as $d\to 1$ -- the blocks become purely transverse! This means that accordingly the bootstrap equations develop a decoupled sector consisting of the $L=0,1$ blocks at $z=\bar z$.  Altogether, these results imply that for a generic four point function the crossing equations take the schematic form in the limit $d\to 1$:
\begin{eqnarray}
\sum_{L=0} \lambda_{\Delta,0}^2 \left(
\begin{tabular}{c}
$F_{\Delta,0}^{\|}$\\
$F_{\Delta,0}^{\perp}$
\end{tabular}
\right)
+
\sum_{L=1} \lambda_{\Delta,1}^2 \left(
\begin{tabular}{c}
$F_{\Delta,0}^{\|}$\\
$F_{\Delta,1}^{\perp}$
\end{tabular}
\right)
+
\sum_{L>0} \hat \lambda_{\Delta,L}^2 \left(
\begin{tabular}{c}
$0$\\
$(d-1)F_{\Delta,L}^{\perp}$
\end{tabular}
\right)=
\left(
\begin{tabular}{c}
$0$\\
$0$
\end{tabular}
\right).
\end{eqnarray}
In the crossing symmetry relations for four identical scalars we must of course have $\lambda^2_{\Delta,L}=0$ for all odd spins $L$. The equations above shows that the crossing equations split into two parts. Firstly, the parallel equations -- denoted with $\|$ -- involve only scalars/spin-1 blocks with $z=\bar z$ and so form a decoupled sector. This decoupled sector is of course simply the purely $d=1$ bootstrap. Hence solutions to crossing symmetry for $d$ arbitrarily close to one have to at least satisfy the same constraints as those in $d=1$. Once these constraints are solved, the spin-0/spin-1 spectrum is completely fixed. Next step is to satisfy the remaining equations, which can be thought of as determining whether an analytic continuation into transverse space -- denoted by $\perp$ -- of this $d=1$ solution can exist. Indeed, since for $d>1$ we switch on the transverse parts of the spin-0 and spin-1 blocks, we will also need to turn on higher spins to cancel those. If we can solve these extra equations, then the analytic continuation exists and will be a smooth function of $d$. If it doesn't, then we are necessarily faced with a discontinuity in the bounds at $d=1$. In other words, the discontinuity can arise because the transverse parts of the $L=0,1$ blocks and also higher spins are by definition only turned on for $d$ strictly greater than one -- this adds a whole new set of crossing constraints which may not have a solution.

\subsection{Results}

\begin{figure}[h]
	\centering
		\includegraphics[width=15cm]{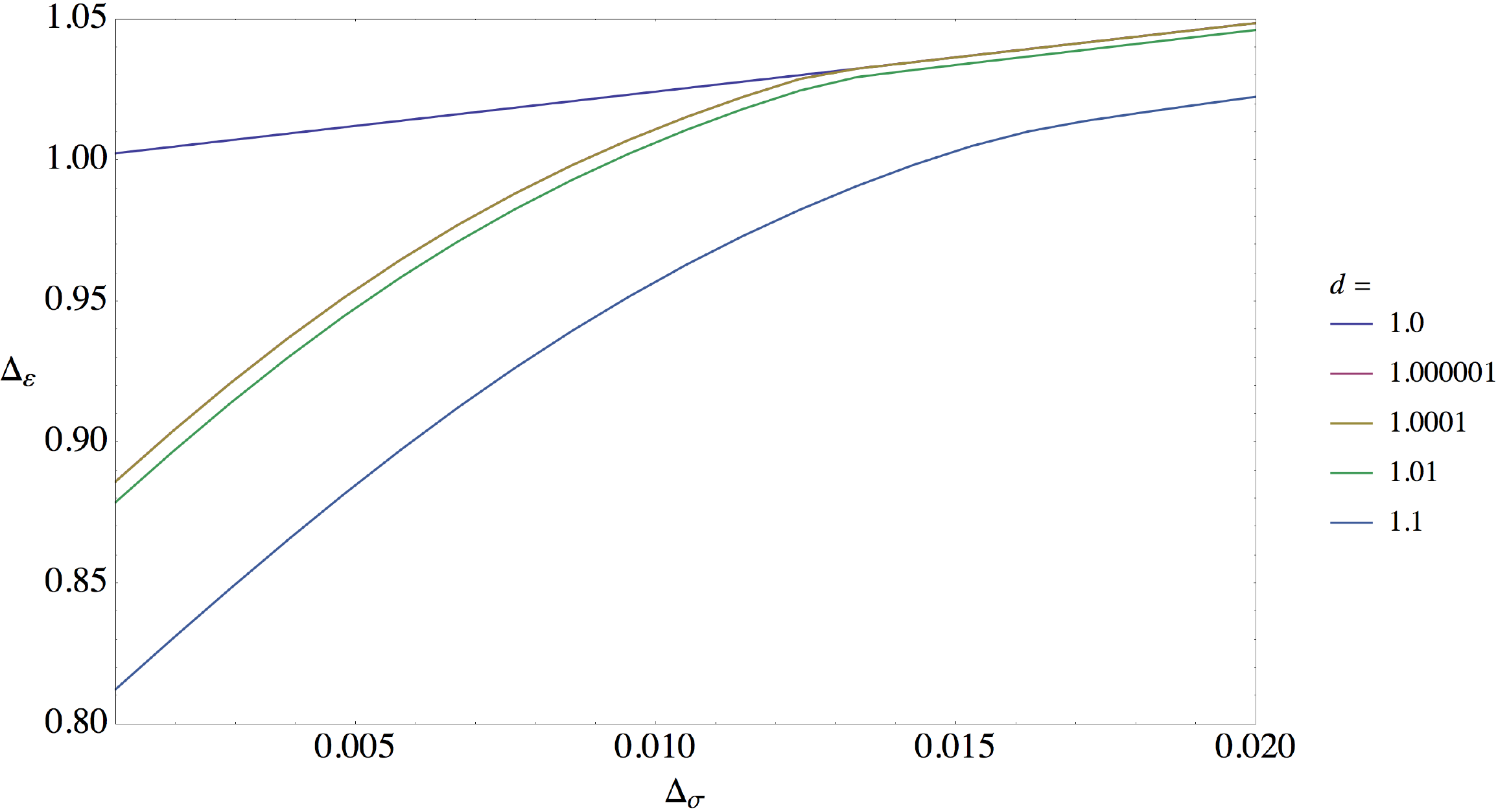}
		\caption{Bounds in $d=1$ and $d=1+\epsilon$. The discontinuity is due to the extra constraints from transverse derivatives which are present for $d>1$. These bounds are evaluated with the constraints truncated at only 10 components ($n_{max} = 3$), but the discontinuity moves further from $d=1$ as the number of components increases.}
	\label{fig:D1Discontinuity}
\end{figure}

To determine if there is a discontinuity, we examine the $d\to 1$ limit numerically. Figure~\ref{fig:D1Discontinuity} shows bounds for dimensions successively closer to $d=1$. Also shown is the strict $d=1$ bound where we keep only the constraints along $z=\bar z$.
These bounds are computed with a small number of derivatives -- 10 components. This is because our bounds can only get stronger with more constraints, so it is sufficient to find a discontinuity in this simple case where we can be sure that our numerics are under control. In the plot, the difference between the $d=1.0001$ and $d=1.000001$ bounds are less than $.0001\%$, indicating that they have stopped progressing towards the $d=1$ bound. Also, the $d=1$ limit seems to exhibit a kink at the point where the bound transitions to $d=1$ curve. As explained in the previous section, in this regime the analytic continuation of the $d=1$ solution into the transverse space exists.

\begin{figure}[htbp]
	\centering
		\includegraphics[width=12cm]{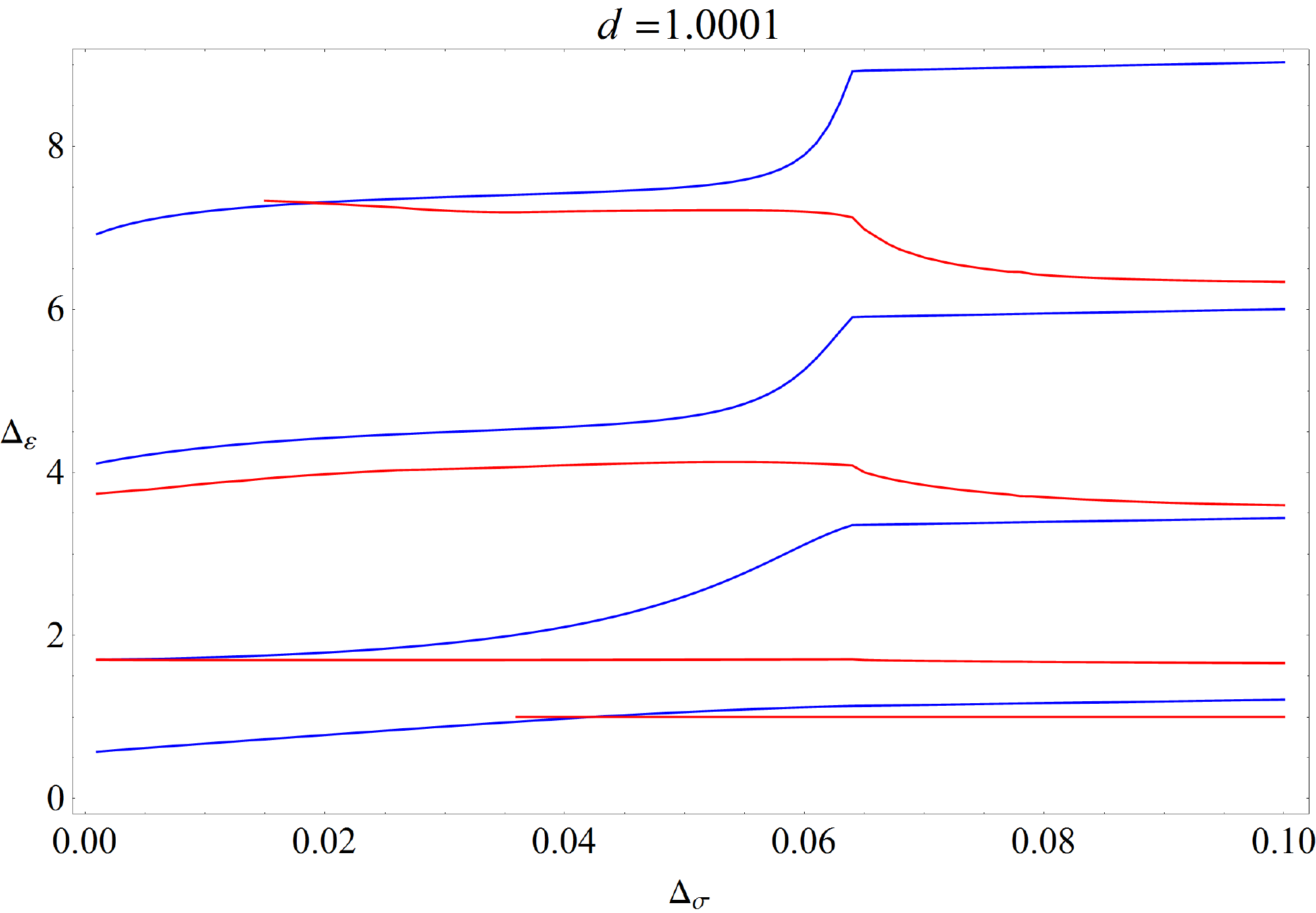}
	\caption{The spin-0 and spin-2 spectrum (blue and red respectively) for $d=1.0001$. Analysis done with 78 components ($n_{max}=11$). The ``plateaus'' which are visible start when the bound equals that for $d=1$.}
	\label{fig:D1spectrum}
\end{figure}

Now that we are confident that there is indeed a discontinuity, we can do a more detailed numerical analysis, and read off the spectrum along the bound.The low lying $L=0$ and $L=2$ operators are shown in figure \ref{fig:D1spectrum}.
Adding derivatives has moved the kink in the bound all the way to $\Delta_{\sigma}\simeq 0.07$. Here we are actually quite confident that the feature corresponds to a true non-analyticity in the limit where we include infinite constraints, since we expect the $d=1$ bound to precisely saturate the straight line $\Delta_{\varepsilon}=1+2\Delta_{\sigma}$. This corresponds to a generalized free fermion \cite{Gaiotto2014}, whose four point function is given by
\begin{eqnarray}
\langle \psi(x_1)\psi(x_2)\psi(x_4)\psi(x_4)\rangle=\frac{\mbox{sgn}(x_1-x_2)\,\mbox{sgn}(x_3-x_4)}{x_{12}^{2\ds}\,x_{34}^{2\ds}}\,\left[1+\left(\frac uv\right)^{\ds}-u^{\ds}\right],  
\label{gff}
\end{eqnarray}
where $d=1$ forces $v=(1-\sqrt u)^2$. This has an expansion in terms of $d=1$ conformal blocks with dimensions $\Delta=1+2\ds+2n$ for $n$ positive integer, something we can see in figure \ref{fig:D1spectrum} for $\Delta_\sigma$ above the kink . This result goes some way in helping us understand the origin of the discontinuity. Indeed, given the correlation function above, the natural guess for its $d>1$ extension is to simply drop the $d=1$ constraint relating the $v$ to the $u$ cross-ratio. But this cannot work, since the expansion of such a four point function will include only odd-spin conformal blocks, whereas we are assuming that the $\sigma$ field is bosonic, and hence the correlation function decomposition should only include even spins. How is it then that nevertheless we obtain (\ref{gff}) in $d=1$? The answer comes from the analysis of the previous section, where we showed that the $z=\bar z$ parts of the scalar and spin-1 blocks actually match in the $d\to 1$ limit. Hence, we may interpret the discontinuity as coming from the fact that $d=1$ is special in allowing us access to fermionic correlation functions using scalar conformal blocks. We expect then that repeating the bootstrap computations for $d>1$ allowing odd instead of even spin conformal blocks we should see a continuous limit -- a result which we have confirmed.

It is interesting to compare these results with models in the literature. The planar interface model of Wallace and Zia discusses the dynamics of a codimension one defect separating two different phases of a $d$-dimensional thermodynamical system \cite{Wallace1979}. The model is a simple DBI action for a $(d-1)$-dimensional brane which is expanded about an infinitely extended surface. One finds there is a weakly coupled UV fixed point for small $\epsilon=d-1$, with critical exponents known up to four loops  \cite{Forster1981}. It will be sufficient for our purposes to quote the leading result
	\begin{eqnarray}
	\nu=\frac{1}{\epsilon}+\mathcal O(1)
	\end{eqnarray}
Recall that $\nu$ is associated with the divergence of the correlation length with as $T\to T_c$. It determines the dimension of $\varepsilon$ in the Ising model, via
\begin{eqnarray}
\Delta_{\varepsilon}=d-\frac{1}\nu\simeq 1+\mathcal O(\epsilon^2)
\end{eqnarray}
Similarly, the droplet model \cite{Bruce1981} considers the configuration energy of surface tension of spherical droplets. Near $d=1$ the droplet distribution function can be computed exactly, and from this the magnetization. At the fixed point we can read off the $\eta$ critical exponent:
	\begin{eqnarray}
	d+\eta-2=\frac{8}{\pi}(\epsilon)^{-1-\epsilon/2} e^{-1-2 C-2/\epsilon}
	\end{eqnarray}
with $C\simeq 0.577$ the Euler constant. The conformal dimension of the field $\sigma$ is then
\begin{eqnarray}
\Delta_{\sigma}=\frac 12 (d-2+\eta)=\frac{4}{\pi}(\epsilon)^{-1-\epsilon/2} e^{-1-2 C-2/\epsilon}
\end{eqnarray}

It is important to note, that {\em a priori} there is no fundamental reason for claiming that these critical exponents are those of the Ising model -- indeed there are arguments that these models do not fully capture the interface free energy of the Ising model \cite{Huse1985} -- so this remains a conjecture. Nevertheless we can take this model as face value and try to find it in our plots. As should be clear however, the discontinuity in our bounds implies that there is no hope of matching the theoretical analysis of the droplet model, since they predict $\ds\to 0$ and $\Delta_{\varepsilon}\to 1$ when $d\to 1$. Clearly all such models are then ruled out as unitary conformally invariant fixed points.

\section{Discussion}\label{sec:discussion}

We have derived bounds on operator dimensions for conformal field theories in $1<d<2$. These results turn out to be surprisingly strong: they completely and clearly rule out a large set of models and predictions for the critical exponents of the Ising model universality class. In particular, the Borel-resummed $\epsilon$-expansion, which compares very favourably with the results of the bootstrap for $2<d<4$, is already ruled out even at $d=1.875$. Overall, there are two sets of approaches for studying Ising-like systems in $1<d<2$: theoretical analyses, such as $\epsilon$-expansion, transfer matrix methods, and the droplet model; and Monte-Carlo simulations of the Ising model on fractal spaces. In every case we have found these models cannot describe unitary CFTs.

Recently the authors of \cite{Hogervorst2014} have noticed that free field theories, and in all likelihood even the Wilson-Fisher fixed point, is non-unitary for {\em any} fractional dimension. Could this be an (admitedly somewhat trivial) explanation for our results? It is hard to say. After all the bootstrap {\em does} work for fractional $d>2$. It would be surprising if it would stop being so for $d\lesssim 2$. The reason it does work for $d>2$ is that such non-unitarity shows up only at relatively high values of the conformal dimension of operators, and so this has a relatively small effect on say the dimension of the first scalar in the $\sigma \times \sigma$ OPE. However such operators are expected to appear at lower dimensions for $d<2$. In any case, we might naively think that continuity would imply that our results should match at least the Borel-resummed $\epsilon$-expansion for $d$ close to two. The fact that they do not suggests something fundamentally new is happening. We have argued that the bootstrap itself gives qualitative evidence for this, through the fact that the single kink present above $d=2$ surprisingly splits into two below it. In the context of $\lambda \phi^4$ theory, we also expect that indeed $d<2$ is different, since the UV of the renormalization group flow down to the hypothetical Wilson-Fisher fixed point is a non-unitary free theory. Accordingly $\phi^4$ acts as a perturbation with negative conformal dimension, which is unusual to say the least. Perhaps this might ultimately be the explanation for our results, but at this point it remains speculative.

Our bounds do not provide much clarity to the complex collection of results surrounding the Ising model on fractal lattices. Of course there is no reason {\em a priori} to expect any of these (non-translationally invariant) theories to lie within our bounds, although they do exhibit critical behaviour. It would be interesting to show that fractal Ising models come closer to our bounds as lacunarity decreases, however there are currently not enough numerical results for a study of this type to be pursued. And, given that the Wilson-Fisher fixed point appears to no longer be a unitary CFT in this regime it is entirely possible that a decrease in lacunarity will have no impact on the connection between fractal Ising models and our bounds. Another avenue for future research would be to consider fractal lattices generated by a random deletion of subsquares, which restores an average translational invariance in the large $k$ limit. Perhaps this, or some other topology, might produce a fractal Ising model corresponding to a unitary CFT in noninteger dimensions, but at this point it does not look too promising. 

Our results for $d\to 1$ are rather interesting. On very general grounds we have excluded models which give $(\Delta_\sigma,\Delta_{\varepsilon})\to (0,1)$ as $d\to 1$. This is precisely the behaviour expected for the Ising model in this limit -- {\em e.g.} the spin field correlation function becomes a constant close to the $T=0$ ``critical point'' in $d=1$. Hence this is perhaps the strongest evidence for our claim that the Wilson-Fisher fixed point is very different below $d<2$. Let us consider the nature of the solution to crossing symmetry obtained by setting ourselves at the bound and for $\Delta_\sigma$ sufficiently large such that we are above the kink. Due to the peculiar $d\to 1$ limit, the scalar sector and the higher spin sector decouple, and furthermore the $z=\bar z$ piece of the solution is given solely in terms of the scalar blocks. This piece is reproduced by the four point function of a generalized free fermion. It would be extremely interesting to determine what the solution looks like for $z\neq \bar z$. This would require understanding the conformal blocks in this limit analytically, which unfortunately seems difficult. Regarding the kink itself, we see that in spite of the very sharp change in the properties of the spectrum, no operator rearrangements are visible. In particular we see no analog of ``null states'' as in previous studies \cite{El-Showk2014}. This suggests that this transition may be kinematical in nature, and not due to the existence of an interesting conformal field theory at this point.

Overall, we may say the results in this note are a double edged sword: on the one hand, they exemplify the power of the conformal bootstrap in cutting through large swaths of conformal theory space and eliminating hypothetical fixed points; but on the other hand they remind us that there may be interesting critical systems which it cannot capture -- if there is indeed a non-unitary Wilson-Fisher fixed point for $d<2$, we will never see it. This is because the bootstrap approach depends on positivity in a crucial way, and this in turn follows from unitarity. In this respect, it would be interesting to pursue the approach promoted by Gliozzi \cite{Gliozzi2014,Gliozzi2013} -- although it is less developed and understood, at least for such systems it might be superior, since it does not depend on unitarity.

\acknowledgments{
We would like to thank S. El-Showk, B. van Rees, V. Rychkov, and A. Vichi for useful discussions and comments. During this work MFP was supported by US DOE-grant DE-SC0010010 and by a Marie Curie Intra-European Fellowship of the
European Community’s 7th Framework Programme under contract number PIEF-GA-2013-623606, while JG is supported by US Department of Energy under contract DE-SC0010010 and a Galkin fellowship.}
\bibliographystyle{JHEP}
\bibliography{paper}

\providecommand{\href}[2]{#2}\begingroup\raggedright\begin{thebibliography}{10}

\bibitem{Ferrara:1974ny}
S.~Ferrara, R.~Gatto, and A.~F. Grillo, {\it {Properties of Partial Wave
  Amplitudes in Conformal Invariant Field Theories}},  {\em Nuovo Cim.} {\bf
  A26} (1975) 226.

\bibitem{Ferrara:1974pt}
S.~Ferrara, R.~Gatto, and A.~F. Grillo, {\it {Positivity Restrictions on
  Anomalous Dimensions}},  {\em Phys. Rev.} {\bf D9} (1974) 3564.

\bibitem{Ferrara:1973yt}
S.~Ferrara, A.~F. Grillo, and R.~Gatto, {\it {Tensor representations of
  conformal algebra and conformally covariant operator product expansion}},
  {\em Annals Phys.} {\bf 76} (1973) 161--188.

\bibitem{Ferrara:1971vh}
S.~Ferrara, A.~F. Grillo, and R.~Gatto, {\it {Manifestly conformal covariant
  operator-product expansion}},  {\em Lett. Nuovo Cim.} {\bf 2S2} (1971)
  1363--1369.

\bibitem{Ferrara:1974nf}
S.~Ferrara, A.~F. Grillo, R.~Gatto, and G.~Parisi, {\it {Analyticity Properties
  and Asymptotic Expansions of Conformal Covariant Green's Functions}},  {\em
  Nuovo Cim.} {\bf A19} (1974) 667--695.

\bibitem{Ferrara:1973vz}
S.~Ferrara, A.~F. Grillo, G.~Parisi, and R.~Gatto, {\it {Covariant expansion of
  the conformal four-point function}},  {\em Nucl. Phys.} {\bf B49} (1972)
  77--98.

\bibitem{Polyakov:1974gs}
A.~M. Polyakov, {\it {Nonhamiltonian approach to conformal quantum field
  theory}},  {\em Zh. Eksp. Teor. Fiz.} {\bf 66} (1974) 23--42.

\bibitem{Rattazzi:2008pe}
R.~Rattazzi, V.~S. Rychkov, E.~Tonni, and A.~Vichi, {\it {Bounding scalar
  operator dimensions in 4D CFT}},  {\em JHEP} {\bf 12} (2008) 031,
  [\href{http://arxiv.org/abs/0807.0004}{{\tt arXiv:0807.0004}}].

\bibitem{Vichi:2011ux}
A.~Vichi, {\it {Improved bounds for CFT's with global symmetries}},  {\em JHEP}
  {\bf 1201} (2012) 162, [\href{http://arxiv.org/abs/1106.4037}{{\tt
  arXiv:1106.4037}}].

\bibitem{Vichi-thesis}
A.~Vichi, {\em {A New Method to Explore Conformal Field Theories in Any
  Dimension}}.
\newblock Ph.D. Thesis, EPFL, 2011, 164 pp.

\bibitem{Rychkov:2009ij}
V.~S. Rychkov and A.~Vichi, {\it {Universal Constraints on Conformal Operator
  Dimensions}},  {\em Phys. Rev.} {\bf D80} (2009) 045006,
  [\href{http://arxiv.org/abs/0905.2211}{{\tt arXiv:0905.2211}}].

\bibitem{Rattazzi:2010gj}
R.~Rattazzi, S.~Rychkov, and A.~Vichi, {\it {Central Charge Bounds in 4D
  Conformal Field Theory}},  {\em Phys. Rev.} {\bf D83} (2011) 046011,
  [\href{http://arxiv.org/abs/1009.2725}{{\tt arXiv:1009.2725}}].

\bibitem{Rattazzi:2010yc}
R.~Rattazzi, S.~Rychkov, and A.~Vichi, {\it {Bounds in 4D Conformal Field
  Theories with Global Symmetry}},  {\em J. Phys.} {\bf A44} (2011) 035402,
  [\href{http://arxiv.org/abs/1009.5985}{{\tt arXiv:1009.5985}}].

\bibitem{Poland:2011ey}
D.~Poland, D.~Simmons-Duffin, and A.~Vichi, {\it {Carving Out the Space of 4D
  CFTs}},  {\em JHEP} {\bf 1205} (2012) 110,
  [\href{http://arxiv.org/abs/1109.5176}{{\tt arXiv:1109.5176}}].

\bibitem{Poland:2010wg}
D.~Poland and D.~Simmons-Duffin, {\it {Bounds on 4D Conformal and
  Superconformal Field Theories}},  {\em JHEP} {\bf 1105} (2011) 017,
  [\href{http://arxiv.org/abs/1009.2087}{{\tt arXiv:1009.2087}}].

\bibitem{Liendo:2012hy}
P.~Liendo, L.~Rastelli, and B.~C. van Rees, {\it {The Bootstrap Program for
  Boundary CFT${}_d$}},  {\em JHEP} {\bf 1307} (2013) 113,
  [\href{http://arxiv.org/abs/1210.4258}{{\tt arXiv:1210.4258}}].

\bibitem{Kos:2013tga}
F.~Kos, D.~Poland, and D.~Simmons-Duffin, {\it {Bootstrapping the $O(N)$ Vector
  Models}},  \href{http://arxiv.org/abs/1307.6856}{{\tt arXiv:1307.6856}}.

\bibitem{Gliozzi2013}
F.~Gliozzi, {\it {More constraining conformal bootstrap}},  {\em
  Phys.Rev.Lett.} {\bf 111} (2013) 161602,
  [\href{http://arxiv.org/abs/1307.3111}{{\tt arXiv:1307.3111}}].

\bibitem{ElShowk:2012ht}
S.~El-Showk, M.~F. Paulos, D.~Poland, S.~Rychkov, D.~Simmons-Duffin, and
  A.~Vichi, {\it {Solving the 3D Ising Model with the Conformal Bootstrap}},
  {\em Phys.Rev.} {\bf D86} (2012) 025022,
  [\href{http://arxiv.org/abs/1203.6064}{{\tt arXiv:1203.6064}}].

\bibitem{El-Showk:2013nia}
S.~El-Showk, M.~F. Paulos, D.~Poland, S.~Rychkov, D.~Simmons-Duffin, and
  A.~Vichi, {\it {Conformal Field Theories in Fractional Dimensions}},  {\em
  Phys. Rev. Lett., to appear} (2014)
  [\href{http://arxiv.org/abs/1309.5089}{{\tt arXiv:1309.5089}}].

\bibitem{ElShowk:2012hu}
S.~El-Showk and M.~F. Paulos, {\it {Bootstrapping Conformal Field Theories with
  the Extremal Functional Method}},  {\em Phys. Rev. Lett.} {\bf 111} (2012)
  241601, [\href{http://arxiv.org/abs/1211.2810}{{\tt arXiv:1211.2810}}].

\bibitem{Beem:2013qxa}
C.~Beem, L.~Rastelli, and B.~C. van Rees, {\it {The $N=4$ Superconformal
  Bootstrap}},  \href{http://arxiv.org/abs/1304.1803}{{\tt arXiv:1304.1803}}.

\bibitem{Gaiotto2014}
D.~Gaiotto, D.~Mazac, and M.~F. Paulos, {\it {Bootstrapping the 3d Ising twist
  defect}},  {\em JHEP} {\bf 1403} (2014) 100,
  [\href{http://arxiv.org/abs/1310.5078}{{\tt arXiv:1310.5078}}].

\bibitem{Fitzpatrick:2012yx}
A.~L. Fitzpatrick, J.~Kaplan, D.~Poland, and D.~Simmons-Duffin, {\it {The
  Analytic Bootstrap and AdS Superhorizon Locality}},  {\em JHEP} {\bf 1312}
  (2013) 004, [\href{http://arxiv.org/abs/1212.3616}{{\tt arXiv:1212.3616}}].

\bibitem{Komargodski:2012ek}
Z.~Komargodski and A.~Zhiboedov, {\it {Convexity and Liberation at Large
  Spin}},  {\em JHEP} {\bf 1311} (2013) 140,
  [\href{http://arxiv.org/abs/1212.4103}{{\tt arXiv:1212.4103}}].

\bibitem{Alday2013}
L.~F. Alday and A.~Bissi, {\it {Higher-spin correlators}},  {\em JHEP} {\bf
  1310} (2013) 202, [\href{http://arxiv.org/abs/1305.4604}{{\tt
  arXiv:1305.4604}}].

\bibitem{Alday2014c}
L.~F. Alday, A.~Bissi, and T.~Lukowski, {\it {Lessons from crossing symmetry at
  large N}},  \href{http://arxiv.org/abs/1410.4717}{{\tt arXiv:1410.4717}}.

\bibitem{Dolan2011}
F.~Dolan and H.~Osborn, {\it Conformal partial waves: further mathematical
  results},  {\em arXiv preprint arXiv:1108.6194} (2011).

\bibitem{Dolan2004}
F.~Dolan and H.~Osborn, {\it Conformal partial waves and the operator product
  expansion},  {\em Nuclear Physics B} {\bf 678} (2004), no.~1 491--507.

\bibitem{Dolan2001}
F.~Dolan and H.~Osborn, {\it Conformal four point functions and the operator
  product expansion},  {\em Nuclear Physics B} {\bf 599} (2001), no.~1
  459--496.

\bibitem{Fitzpatrick:2013sya}
A.~L. Fitzpatrick, J.~Kaplan, and D.~Poland, {\it {Conformal Blocks in the
  Large D Limit}},  {\em JHEP} {\bf 1308} (2013) 107,
  [\href{http://arxiv.org/abs/1305.0004}{{\tt arXiv:1305.0004}}].

\bibitem{Costa:2011dw}
M.~S. Costa, J.~Penedones, D.~Poland, and S.~Rychkov, {\it {Spinning Conformal
  Blocks}},  {\em JHEP} {\bf 1111} (2011) 154,
  [\href{http://arxiv.org/abs/1109.6321}{{\tt arXiv:1109.6321}}].

\bibitem{Hogervorst:2013sma}
M.~Hogervorst and S.~Rychkov, {\it {Radial Coordinates for Conformal Blocks}},
  {\em Phys.Rev.} {\bf D87} (2013) 106004,
  [\href{http://arxiv.org/abs/1303.1111}{{\tt arXiv:1303.1111}}].

\bibitem{Hogervorst:2013kva}
M.~Hogervorst, H.~Osborn, and S.~Rychkov, {\it {Diagonal Limit for Conformal
  Blocks in $d$ Dimensions}},  {\em JHEP} {\bf 1308} (2013) 014,
  [\href{http://arxiv.org/abs/1305.1321}{{\tt arXiv:1305.1321}}].

\bibitem{El-Showk2014}
S.~El-Showk, M.~F. Paulos, D.~Poland, S.~Rychkov, D.~Simmons-Duffin, {\em
  et~al.}, {\it {Solving the 3d Ising Model with the Conformal Bootstrap II.
  c-Minimization and Precise Critical Exponents}},
  \href{http://arxiv.org/abs/1403.4545}{{\tt arXiv:1403.4545}}.

\bibitem{Pelissetto2002}
A.~Pelissetto and E.~Vicari, {\it {Critical phenomena and renormalization group
  theory}},  {\em Phys.Rept.} {\bf 368} (2002) 549--727,
  [\href{http://arxiv.org/abs/cond-mat/0012164}{{\tt cond-mat/0012164}}].

\bibitem{Wilson:1971dc}
K.~G. Wilson and M.~E. Fisher, {\it {Critical exponents in 3.99 dimensions}},
  {\em Phys. Rev. Lett.} {\bf 28} (1972) 240--243.

\bibitem{Wilson:1972cf}
K.~G. Wilson, {\it {Quantum field theory models in less than four-dimensions}},
   {\em Phys.Rev.} {\bf D7} (1973) 2911--2926.

\bibitem{Wilson:1973jj}
K.~Wilson and J.~B. Kogut, {\it {The Renormalization group and the epsilon
  expansion}},  {\em Phys.Rept.} {\bf 12} (1974) 75--200.

\bibitem{LeGuillou1985}
J.~Le~Guillou and J.~Zinn-Justin, {\it Accurate critical exponents from the
  $\varepsilon$-expansion},  {\em Journal de Physique Lettres} {\bf 46} (1985),
  no.~4 137--141.

\bibitem{LeGuillou1980}
J.~Le~Guillou and J.~Zinn-Justin, {\it {Critical Exponents from Field Theory}},
   {\em Phys.Rev.} {\bf B21} (1980) 3976--3998.

\bibitem{Holovatch1993}
Y.~Holovatch, {\it {Critical exponents of Ising like systems in general
  dimensions}},  {\em Theor.Math.Phys.} {\bf 96} (1993) 1099--1110.

\bibitem{Bonnier1991}
B.~Bonnier and M.~Hontebeyrie, {\it Critical properties of the d-dimensional
  ising model from a variational method},  {\em Journal de Physique I} {\bf 1}
  (1991), no.~3 331--338.

\bibitem{Novotny1992}
M.~Novotny, {\it Critical exponents for the ising model between one and two
  dimensions},  {\em Physical Review B} {\bf 46} (1992), no.~5 2939.

\bibitem{Wallace1979}
D.~Wallace and R.~Zia, {\it {THE EUCLIDEAN GROUP AS A DYNAMICAL SYMMETRY OF
  SURFACE FLUCTUATIONS: THE PLANAR INTERFACE AND CRITICAL BEHAVIOR}},  {\em
  Phys.Rev.Lett.} {\bf 43} (1979) 808.

\bibitem{Bruce1981}
A.~Bruce and D.~Wallace, {\it Droplet theory of low-dimensional ising models},
  {\em Physical Review Letters} {\bf 47} (1981), no.~24 1743.

\bibitem{monceau1998magnetic}
P.~Monceau, M.~Perreau, and F.~H{\'e}bert, {\it Magnetic critical behavior of
  the ising model on fractal structures},  {\em Physical Review B} {\bf 58}
  (1998), no.~10 6386.

\bibitem{Gefen1984}
Y.~Gefen, A.~Aharony, and B.~B. Mandelbrot, {\it Phase transitions on fractals.
  iii. infinitely ramified lattices},  {\em Journal of Physics A: Mathematical
  and General} {\bf 17} (1984), no.~6 1277.

\bibitem{Gefen1984a}
Y.~Gefen, A.~Aharony, Y.~Shapir, and B.~B. Mandelbrot, {\it Phase transitions
  on fractals. ii. sierpinski gaskets},  {\em Journal of Physics A:
  Mathematical and General} {\bf 17} (1984), no.~2 435.

\bibitem{Gefen1980}
Y.~Gefen, B.~B. Mandelbrot, and A.~Aharony, {\it Critical phenomena on fractal
  lattices},  {\em Physical Review Letters} {\bf 45} (1980), no.~11 855.

\bibitem{Pappadopulo:2012jk}
D.~Pappadopulo, S.~Rychkov, J.~Espin, and R.~Rattazzi, {\it {OPE Convergence in
  Conformal Field Theory}},  {\em Phys.Rev.} {\bf D86} (2012) 105043,
  [\href{http://arxiv.org/abs/1208.6449}{{\tt arXiv:1208.6449}}].

\bibitem{dantzig1955generalized}
G.~B. Dantzig, A.~Orden, P.~Wolfe, {\em et~al.}, {\it The generalized simplex
  method for minimizing a linear form under linear inequality restraints},
  {\em Pacific Journal of Mathematics} {\bf 5} (1955), no.~2 183--195.

\bibitem{Wilson1972a}
K.~G. Wilson and M.~E. Fisher, {\it {Critical exponents in 3.99 dimensions}},
  {\em Phys.Rev.Lett.} {\bf 28} (1972) 240--243.

\bibitem{LeGuillou1987}
J.~Le~Guillou and J.~Zinn-Justin, {\it {Accurate critical exponents for Ising
  like systems in noninteger dimensions}}, .

\bibitem{Parisi1993}
G.~Parisi, {\it {Field theoretic approach to second order phase transitions in
  two-dimensional and three-dimensional systems}}, .

\bibitem{LeGuillou1977}
J.~Le~Guillou and J.~Zinn-Justin, {\it {Critical Exponents for the N Vector
  Model in Three-Dimensions from Field Theory}},  {\em Phys.Rev.Lett.} {\bf 39}
  (1977) 95--98.

\bibitem{Bab2005}
M.~Bab, G.~Fabricius, and E.~Albano, {\it Critical behavior of an ising system
  on the sierpinski carpet: A short-time dynamics study},  {\em Physical Review
  E} {\bf 71} (2005), no.~3 036139.

\bibitem{Bab2009}
M.~Bab, G.~Fabricius, and E.~Albano, {\it Critical exponents of the ising model
  on low-dimensional fractal media},  {\em Physica A: Statistical Mechanics and
  its Applications} {\bf 388} (2009), no.~4 370--378.

\bibitem{Bonnier1986}
B.~Bonnier, Y.~Leroyer, and C.~Meyers, {\it {CRITICAL EXPONENTS FOR ISING LIKE
  SYSTEMS ON SIERPINSKI CARPETS}}, .

\bibitem{Bonnier1987a}
B.~Bonnier, Y.~Leroyer, and C.~Meyers, {\it {REAL SPACE RENORMALIZATION GROUP
  STUDY OF FRACTAL ISING MODELS}}, .

\bibitem{Bonnier1989}
B.~Bonnier, Y.~Leroyer, and C.~Meyers, {\it {HIGH TEMPERATURE EXPANSIONS ON
  SIERPINSKI CARPETS}},  {\em Phys. Rev. B} (1989).

\bibitem{Carmona1998}
J.~Carmona, J.~Ruiz-Lorenzo, U.~Marconi, and A.~Taranc{\'o}n, {\it Phase
  transitions on sierpinski fractals},  {\em Phys. Rev. B} {\bf 58} (1998),
  no.~cond-mat/9802018 14387.

\bibitem{Gefen1983}
Y.~Gefen, Y.~Meir, B.~B. Mandelbrot, and A.~Aharony, {\it Geometric
  implementation of hypercubic lattices with noninteger dimensionality by use
  of low lacunarity fractal lattices},  {\em Physical Review Letters} {\bf 50}
  (1983), no.~3 145.

\bibitem{PhysRevE.74.041123}
M.~A. Bab, G.~Fabricius, and E.~V. Albano, {\it Discrete scale invariance
  effects in the nonequilibrium critical behavior of the ising magnet on a
  fractal substrate},  {\em Phys. Rev. E} {\bf 74} (Oct, 2006) 041123.

\bibitem{Hogervorst2013a}
M.~Hogervorst, H.~Osborn, and S.~Rychkov, {\it {Diagonal Limit for Conformal
  Blocks in $d$ Dimensions}},  {\em JHEP} {\bf 1308} (2013) 014,
  [\href{http://arxiv.org/abs/1305.1321}{{\tt arXiv:1305.1321}}].

\bibitem{Forster1981}
D.~Forster and A.~Gabriunas, {\it Critical behavior of an
  $\varepsilon$-dimensional planar interface},  {\em Physical Review A} {\bf
  24} (1981) 598--600.

\bibitem{Huse1985}
D.~A. Huse, W.~van Saarloos, and J.~D. Weeks, {\it Interface hamiltonians and
  bulk critical behavior},  {\em Physical Review B} {\bf 32} (1985), no.~1 233.

\bibitem{Hogervorst2014}
M.~Hogervorst, S.~Rychkov, and B.~C. van Rees, {\it {A Cheap Alternative to the
  Lattice?}},  \href{http://arxiv.org/abs/1409.1581}{{\tt arXiv:1409.1581}}.

\bibitem{Gliozzi2014}
F.~Gliozzi and A.~Rago, {\it {Critical exponents of the 3d Ising and related
  models from Conformal Bootstrap}},  {\em JHEP} {\bf 1410} (2014) 42,
  [\href{http://arxiv.org/abs/1403.6003}{{\tt arXiv:1403.6003}}].

\end{thebibliography}\endgroup

\end{document}